\documentclass[preprint,12pt]{elsarticle}
\usepackage[ruled,vlined]{algorithm2e}
\usepackage{multirow}
\usepackage{graphicx}
\usepackage{amssymb,amsmath}
\usepackage{amsthm}

\newtheorem{definition}{Definition}

\newtheorem{property}{Property}

\journal{Pattern Recognition Letters}

\begin{document}

\begin{frontmatter}

%% Title, authors and addresses

%% use the tnoteref command within \title for footnotes;
%% use the tnotetext command for the associated footnote;
%% use the fnref command within \author or \address for footnotes;
%% use the fntext command for the associated footnote;
%% use the corref command within \author for corresponding author footnotes;
%% use the cortext command for the associated footnote;
%% use the ead command for the email address,
%% and the form \ead[url] for the home page:
%%
%% \title{Title\tnoteref{label1}}
%% \tnotetext[label1]{}
%% \author{Name\corref{cor1}\fnref{label2}}
%% \ead{email address}
%% \ead[url]{home page}
%% \fntext[label2]{}
%% \cortext[cor1]{}
%% \address{Address\fnref{label3}}
%% \fntext[label3]{}

\title{Towards an Efficient Discovery of the Topological Representative Subgraphs}

%% use optional labels to link authors explicitly to addresses:
\address[1]{LIMOS - Blaise Pascal University - Clermont University, Clermont-Ferrand 63000, France.}
\address[2]{LIMOS - CNRS UMR 6158, Aubi\`ere 63173, France.\\}
\address[3]{Department of Computer Science - FSEGJ - University of Jendouba, UMA Street, Jendouba 8100, Tunisia.}
\address[4]{European Bioinformatics Institute, Hinxton, Cambridge, CB10 1SD, United Kingdom.}

\author[1,2]{Wajdi Dhifli}
\author[3]{Mohamed Moussaoui}
\author[4]{Rabie Saidi}
\author[1,2]{Engelbert Mephu Nguifo\footnote{Email: mephu@isima.fr		Phone: +33 473 407 629		Fax: +33 473 407 639}}

%\titlenote{Dr.~Trovato insisted his name be first.}\\
%       \affaddr{Institute for Clarity in Documentation}\\
%       \affaddr{1932 Wallamaloo Lane}\\
%       \affaddr{Wallamaloo, New Zealand}\\
%       \email{trovato@corporation.com}
%
%\author{}
%
%\address{}

\begin{abstract}
With the emergence of graph databases, the task of frequent subgraph discovery has been extensively addressed. Although the proposed approaches in the literature have made this task feasible, the number of discovered frequent subgraphs is still very high to be efficiently used in any further exploration. Feature selection based on exact or approximate structural similarity is a way to reduce the high number of frequent subgraphs. However, current structural similarity strategies are not efficient enough in many real-world applications, besides, the combinatorial nature of graphs makes it computationally very costly. In order to select a smaller yet structurally irredundant set of subgraphs, we propose a novel approach that mines the top-k topological representative subgraphs among the frequent ones. Our approach allows detecting hidden structural similarities that existing approaches are unable to detect such as the density or the diameter of the subgraph. In addition, it can be easily extended using any user defined structural or topological attributes depending on the sought properties. Empirical studies on real and synthetic graph datasets show that our approach is fast and scalable.
\\\end{abstract}

\begin{keyword}
% keywords here, in the form: keyword \sep keyword

%% MSC codes here, in the form: \MSC code \sep code
% or \MSC[2008] code \sep code (2000 is the default)
Feature selection \sep topological representative subgraphs \sep frequent subgraphs \sep graph databases
\end{keyword}

\end{frontmatter}

%%
%% Start line numbering here if you want
%%
% \linenumbers

%% main text
\section{Introduction}
Graphs are one of the most powerful structure to model complex data. In fact, any data composed of entities having relationships, can be represented by a graph where the entities will be seen as the graph nodes and the relationships as the graph edges. With the emergence of graph databases, the task of frequent subgraph discovery has been extensively addressed. Many approaches have been proposed in the literature allowing the extraction of frequent subgraphs in an efficient way. Yet, the number of discovered frequent subgraphs is extremely high which may hinder or even makes unfeasible further exploration.

Feature selection for graph data is a way to tackle the dimensionality problem when the number of frequent subgraphs is very high. As structural similarity represents one major cause of redundancy in frequent subgraphs, many works have been proposed for subgraph selection based on exact or approximate structural similarity \cite{Yan_2003, Thomas_2006, Chaoji_2008, Chen_2008}. Two pioneer works that fall in this type are \cite{Yan_2003} and \cite{Thomas_2006}. If a graph is contained in another one then the small graph is called the subgraph and the big one is called the super graph. In \cite{Yan_2003}, a frequent subgraph is said to be closed if there exists no frequent supergraph having the same support. In \cite{Thomas_2006}, a frequent subgraph is said to be maximal if it has no frequent supergraph. In both cases, only the closed or maximal subgraphs are maintained and the rest of frequent subgraphs are removed. Many works have been proposed based on closed and maximal subgraphs such as \cite{Takigawa_2011, Li_2007}. Although the set of closed or maximal subgraphs is much smaller than the set of frequent ones, the number of subgraphs is still very high in real-world cases. Another fresh work for subgraph selection based on exact isomorphism is \cite{Dhifli_2013}. In this work, authors tried to select the called representative unsubstituted subgraphs using a similarity function that exploits the prior domain knowledge. A fundamental constraint in their selection is that only structurally isomorphic subgraphs are considered for substitution. Although they highly decrease the number of subgraphs, in real-world cases the data is often noisy and very similar subgraphs sometimes slightly differ in structure. Exact structural isomorphism does not help to overcome this issue. Many works have been proposed for subgraph selection based on approximate structural similarity. In \cite{Chaoji_2008}, authors proposed an approach for subgraphs extraction and selection. For selection, the structural similarity between two subgraphs is measured by how much does their maximum common subgraph \cite{Abu_2007} represents from their overall structure. A very close work is \cite{Chen_2008}, where authors proposed an approach for mining a set of structural representative subgraphs among the frequent ones. They adopted a two step based-approach that is based on approximate structural similarity on micro and macro sides. In the first step, they consider a tolerance threshold to summarize approximately isomorphic subgraphs into one representative. In the second step, they collapse multiple structurally similar subgraphs into one representative using a clustering algorithm.

Existing approaches look into every single detail and test the structural similarity of subgraphs by establishing a matching between them. We believe that following this similarity detection strategy is not efficient enough in many real-world applications. On one hand, because the combinatorial nature of graphs makes looking for a possible matching between every pair of subgraphs computationally very costly. On the other hand, exact and even approximate structural similarity are not efficient enough to detect all similar subgraphs in real-world data. Indeed, exact structural similarity does not allow detecting similar yet slightly different subgraphs, and approximate structural similarity has the problem of threshold setting. Since a tight threshold will prevent detecting many similar subgraphs that slightly differ in structure beyond the tolerance threshold and thus preserve a high number of subgraphs. In contrast, a loose threshold will hinder the soundness of the selection because of false positives. This arises the need for a different way to consider the structural similarity such that both close and distant structural similarities would be detected with respect to the soundness of results. Considering topological properties instead of exact or approximate structural isomorphism was inspired by works like \cite{Li_2012, Ranu_2012, Tong_2012, Gibert_2012, Veeramalai_2008, Leskovec_2005, Rodenacker_1990} where authors showed the importance and efficiency of topological attributes in describing graph data. For instance, in \cite{Li_2012}, authors proposed a classification framework based on the assumption that graphs belonging to the same class have similar topological descriptions. Our approach is based on similar assumption and consider that structurally similar subgraphs should have similar topological properties such that even a slight difference does not affect the overall topological similarity. Besides, depending on the application context, a user may be interested only in some specific structural properties. However, considering exact or approximate structural similarity approaches does not allow this specificity. 

In order to overcome these drawbacks and to select a small yet structurally irredundant set of subgraphs, we propose a novel approach that mines the top-k topological representative subgraphs among the frequent ones. At a glance, our approach involves two steps. In the first step, each subgraph is encoded into a topological description-vector containing the corresponding values for a set of topological attributes. In the second step, subgraphs with similar topological descriptions are clustered together and the central subgraph in each cluster is considered as the representative delegate. Our approach overcome the costly isomorphism needed to perform the exact or approximate structural similarity and allows detecting hidden similarities like spectral radius or closeness centrality, that exact or approximate structural similarity approaches are unable to detect. Besides, our approach can be easily extended by enabling the user to target a specific set of topological attributes depending on how important each one is to the application.

The remainder of the paper is organized as follows. Section \ref{sec:methods} presents and defines the preliminary concepts as well as the main algorithm of our approach. In Section \ref{sec:experimental} describes the datasets and the experimental settings. In Section \ref{sec:results}, we present the obtained results and the discussion.

\section{Material and methods}\label{sec:methods}
In this section, we present the fundamental definitions and the formal problem statement of the na\"ive approach for top-k representative subgraph selection and the proposed approach for top-k topological representative subgraph selection. 
\subsection{Preliminaries}
\begin{definition}(Graphs and graph databases)
Let $\mathcal{G}$ be a database of connected graphs. Each graph $G=(V,E,L)$ of $\mathcal{G}$ is given as a collection of nodes $V$ and edges $E$. The nodes of $V$ are labeled within an alphabet $L$. We denote by $|V|$ the number of nodes and by $|E|$ the number of edges.
\end{definition}
\begin{definition}(Subgraph isomorphism)
A labeled graph $G$ is subgraph of another labeled graph $G'$, denoted by $G\subseteq G'$, if there exists an injective function $f: V(G)\rightarrow V(G')$, such that:
\begin{itemize}
\item[-] $\forall v\in V(G), L(v) = L'(f(v))$
\item[-] $\forall(u, v)\in E(G), (f(u),f(v))\in E(G')$ and $L(u, v) = L'(f(u),f(v))$
\end{itemize}
where $L$ and $L'$ are respectively the label functions of $G$ and $G'$. Under these conditions, the function $f$ is called an embedding of $G$ in $G'$, $G$ is called a subgraph of $G'$, and $G'$ is called a supergraph of $G$.
\end{definition}
\begin{definition}(Frequent subgraph)
Given a subgraph $g$, a graph database $\mathcal{G}$, and a minimum frequency threshold $\tau$, let $\mathcal{G}_g$ be the set of graphs where $g$ appears ($i.e.$ $g$ has a subgraph isomorphism in each graph in $\mathcal{G}_g$), such that $\mathcal{G}_g\in \mathcal{G}$. The subgraph $g$ is considered as frequent if $|\mathcal{G}_g|\geq\tau$.
\end{definition}

\textbf{Problem Statement}: Even though the existing approaches for subgraph selection greatly enhanced the selection process, the number of selected subgraphs is still high. Yet, we want to show as few subgraphs as possible so that the user's reviewing efforts are minimized. The general framework of our selection strategy is as follows. Given a set of frequent subgraphs $\Omega$ and an integer $k \in [1 .. |\Omega|]$, we want to select up to $k$ representative subgraphs $\Omega_k\subseteq\Omega$ such that each frequent subgraph $g\in\Omega$ has one representative subgraph-delegate $g'\in\Omega_k$, and each representative subgraph is the closest one to all the subgraph it represents. To do so, the set of frequent subgraphs is divided into $k$ clusters using a clustering algorithm, then the clusters centroids are selected to be the representative subgraph-delegates such that each centroid is representative for all subgraphs within the same cluster.

\subsection{Na\"ive approach}
As we are attempting to select top-k representative subgraphs based on clustering, a fundamental part in our selection framework is the graph encoding which consists in the transformation of each subgraph into a different format that is accepted by the clustering algorithm. A na\"ive solution is to transform the input subgraphs into a context-matrix where each subgraph is represented by a binary vector denoting by 1 or 0 the presence or the absence of the subgraph in each graph in the database. After that, the context-matrix is considered as input for clustering (see Algorithm \ref{alg:naive_algo}).

\begin{algorithm}
\label{alg:naive_algo}
\caption{\textsc{Na\"ive approach}}
%\DontPrintSemicolon
\SetAlgoLined
\LinesNumbered
\KwData{Frequent subgraphs $\Omega$, number of representatives $k$}
\KwResult{Representative subgraphs $\Omega^*$ = \{$g_1, g_2, ..., g_k$\}}
\Begin{
$M \leftarrow \cup_{i=1}^{|\Omega|} V_i$: each subgraph $g\in\Omega$ is encoded into a binary vector $V$ denoting by 1 or 0 correspondingly the presence or the absence of the subgraph in each graph in the database\;
$\Omega^* \leftarrow $Clustering$(M, k)$\;
}
\end{algorithm}
\subsection{Topological representative subgraph selection}
The main idea of our approach is based on the assumption that structurally similar subgraphs should have similar topological properties such that even a slight difference in the structure does not affect the overall similarity. Accordingly, we adopt a two-step selection framework, where in the first step we encode each subgraph into a topological description-vector containing the corresponding values for a set of topological attributes. In the second step, we perform a clustering using the topological description-vectors in order to select one representative subgraph delegate from each set of topologically similar subgraphs.

\subsubsection{Topological attributes}
In the first step of our approach each subgraph is encoded into a topological description-vector. We select a set of topological attributes from the literature \cite{Li_2012, Leskovec_2005} that are interesting and efficient in describing connected graphs. In the following, we list and define the considered attributes:
\begin{enumerate}
\item \textbf{Number of nodes}: The total number of nodes in the graph, also called the graph order $|V|$.
\item \textbf{Number of edges}: The total number of edges in the graph, also called the graph size $|E|$.
\item \textbf{Average degree}: The degree of a node $u$, denoted $deg(u)$, represents the number of nodes adjacent to $u$. The average degree of a graph $G$ is the average value of the degrees of all nodes in $G$. Formally: $ deg(G) = \frac{1}{n} \sum^n_{i=1} deg(u_i)$ where $deg(u_i )$ is the degree of the node $u_i$ and $n$ is the number of nodes in $G$. 
\item \textbf{Density}: The density of a graph $G=(V, E)$ measures how many edges are in $E$ compared to the maximum possible number of edges between the nodes in $V$. Formally: $ den(G) = \frac{2 \mid E\mid}{(\mid V\mid\ast (\mid V\mid -1))}$.
\item \textbf{Average clustering coefficient}: The clustering coefficient of a node $u$, denoted by $c(u)$, measures how complete the neighborhood of $u$ is $i.e.$ $c(u)= \frac{2 e_u}{k_u (k_u - 1))}$ where $k_u$ is the number of neighbors of $u$ and $e_u$ is the number of connected pairs of neighbors. If all the neighbor nodes of u are connected, then the neighborhood of $u$ is complete and we have a clustering coefficient of 1. If no nodes in the neighborhood of $u$ are connected, then the clustering coefficient is 0. The average clustering coefficient of an entire graph $G$ having $n$ nodes, is given as the average value over all the nodes in $G$. Formally: $C(G)= \dfrac{1}{n} \sum_{i=1}^n c(u_i)$.
\item \textbf{Average effective eccentricity}: For a node $u$, the effective eccentricity represents the maximum length of the shortest paths between $u$ and every other node $v$ in $G$, $i.e.$, $e(u) = max\{d(u,v) : v\in V\}$. If $u$ is isolated then $e(u) = 0$. The average effective eccentricity is defined as $Ae(G)= \frac{1}{n}\sum_{i=1}^n e(u_i)$, where $n$ is the number of nodes of $G$.
\item \textbf{Effective diameter}: The effective diameter represents the maximum value of effective eccentricity over all nodes in the graph $G$, $i.e.$, $diam(G) = max\lbrace e(u)\mid u\in V\rbrace$ where $e(u)$ represents the effective eccentricity of $u$ as defined above.
\item \textbf{Effective radius}: The effective radius represents the minimum value of effective eccentricity over all nodes in the graph $G$, $i.e.$, $rad(G) = min\lbrace e(u)\mid u\in V\rbrace$ where $e(u)$ represents the effective eccentricity of $u$.
\item \textbf{Closeness centrality}: The closeness centrality measures how fast information spreads from a given node to other reachable nodes in the graph. For a node $u$, it represents the reciprocal of the average shortest path length between $u$ and every other reachable node in the graph, $i.e.$, $C_c(u) = \frac{n-1}{\sum_{v\in \lbrace V\setminus u\rbrace} d(u,v)}$ where $d(u,v)$ is the length of the shortest path between the nodes $u$ and $v$. For a graph $G$, we consider the average value of closeness centrality of all the nodes, $i.e.$, $C_c(G) = \frac{1}{n} \sum_{i=1}^n u_i$.
\item \textbf{Percentage of central nodes}: Here, we compute the ratio of the number of central nodes from the number of nodes in the graph. A node $u$ is considered as central point if the value of its eccentricity is equal to the effective radius of the graph, $i.e.$, $e(u) = rad(G)$.
\item \textbf{Percentage of end points}: It represents the ratio of the number of end points from the total number of nodes of the graph. A node $u$ is considered as end point if $deg(u) = 1$.
\item \textbf{Number of distinct eigenvalues}: Any graph $G$ can be represented by an adjacency matrix $A$. As the adjacency matrix $A$ has a set of eigenvalues, these eigenvalues are not necessarily different. Here, we count the number of distinct eigenvalues of $A$.
\item \textbf{Spectral radius}: Let $A$ be the adjacency matrix of the graph $G$ and $\leftthreetimes_1 , \leftthreetimes_2 , ..., \leftthreetimes_m$ be the set of eigenvalues of $A$. The spectral radius of $G$, denoted $\rho(G)$, represents the largest magnitude eigenvalue, $i.e.$, $\rho(G) = max(\mid\leftthreetimes_i\mid)$ where $i\in \lbrace 1, .., m\rbrace$.
\item \textbf{Second largest eigenvalue}: The value of the second largest eigenvalue of the adjacency matrix of the graph.
\item \textbf{Energy}: The energy of an adjacency matrix $A$ of a graph $G$ is defined as the squared sum of the eigenvalues of $A$. Formally: $E(G) = \sum^n_{i=1}\leftthreetimes_i^2$.
\item \textbf{Neighborhood impurity}: The impurity degree of a node $u$ belonging to a graph $G$, having a label $L(u)$ and a neighborhood (adjacent nodes) $N(u)$, is defined as $ImpurityDeg(u) = \mid L(v): v \in N(u), L(u)\neq L(v)\mid$. The neighborhood impurity of a graph $G$ represents the average impurity degree over all nodes with positive impurity.
\item \textbf{Link impurity}: An edge $(u, v)$ is considered to be impure if $L(u)\neq L(v)$. The link impurity of a graph $G$ with $k$ edges is defined as: $\frac{\mid(u, v)\in E: L(u)\neq L(v)\mid}{k}$.
\end{enumerate}
As efficiency and scalability remain big challenges for graph mining algorithms, the proposed description is unified which helps to overcome both challenges. On one hand, these attributes present an efficient description that is able to reveal hidden topological similarities that exact and approximate structural isomorphism do not consider. On the other hand, considering a fixed number of descriptors guarantee that the encoded vectors would be of a fixed size no matter what the number of graphs in the database is. Oppositely, the context-vectors in the the na\"ive approach are as big as the number of graphs in the database which is usually very high in real-world applications. This highly affects the scalablity and computational consumption of the selection.

It is also worth mentioning that the topological attributes are not limited to the ones chosen in this work but can be extended by removing or adding other attributes depending on the data and the application goals. For instance, for graphs containing many loops, it could be interesting to consider the number of loops in each subgraph as feature.

\subsubsection{K-medoids clustering}
Here, we discuss the second part of our selection approach which is the clustering step. We use \textit{k-medoids}\cite{Ng_2002} which is a well known clustering algorithm that is widely used in unsupervised learning\cite{Jain_2010}. It takes as input a set of objects $\Omega$ and a number of clusters $k$, and gives as output the $k$ cluster centers (called \textit{medoids}) that we consider as the representative objects. To do so, k-medoids needs these definitions. 

\begin{definition}(Pairwise distance between objects)
Given two objects $O_1$ and $O_2$ correspondingly described by the vectors $X$ and $Y$, the distance between them, denoted $d(O_1,O_2)$, is defined as follows:
\begin{center}
$d(O_1,O_2) = \sum_{i=1}^{|X|} |x_i-y_i|, x_i\in X, y_i\in Y$.
\end{center}
\end{definition}
\begin{definition}(Global distance between objects)
Given a set of objects $\Omega$, the total distance between an object $O$ and all the other ones in $\Omega$ is defined by:
\begin{center}
$D_O = \sum_{i=1}^{|\Omega|-1} d(O - O_i), \forall O_i \in \Omega\setminus O$.
\end{center}
\end{definition}
\begin{definition}(Cluster medoid)
An object $O$ is said to be cluster's medoid $O^*$ (the most centrally located object of the cluster), if it has the minimum sum of distances to all the other objects $O_i$ within the cluster $C$. Formally:
\begin{center}
$D_{O^*} = min_{i=1}^{|C|}(D_{O_i}), \forall O_i\in C$.
\end{center}
The medoid object is considered as representative for all the objects within the same cluster.
\end{definition}

%K-medoids is used to search for $k$ representative objects among the data. 
The general algorithm of k-medoids is described in Algorithm \ref{alg:kmedoids}. First, it starts by randomly selecting $k$ objects from $\Omega$ to be the medoids, $i.e.$, $\Omega^*$. Then, it assigns each non-selected object to the cluster of the nearest medoid. After that, it swaps the $k$ medoid objects with other non-medoid objects aiming to minimize the overall distance. $D(\Omega^*)$ is the total distance before the swap and $D(\Omega'_k)$ is the total distance after the swap. If the cost of the swap $(C = D(\Omega'_k) - D(\Omega^*))$ is strictly negative then the swap is considered as beneficial, otherwise it is ignored. The assignment and swap steps are iteratively performed until no change. Many implementations of k-medoids have been proposed in the literature. PAM \cite{Kaufman_1990} is a pioneer implementation of k-medoids. Later, two other implementations have been proposed which are CLARAN \cite{Kaufman_1990} and CLARANS \cite{Ng_2002}. The main difference between these implementations is in the way of performing the swap where in attempt to make the algorithm more scalable to larger amounts of data. In this work, we use CLARANS since it was shown \cite{Ng_2002} that it is the most efficient implementation for large-scale data clustering and gives similar clustering quality to PAM and CLARAN.

\begin{algorithm}
\label{alg:kmedoids}
\caption{\textsc{K-medoids}}
%\DontPrintSemicolon
\SetAlgoLined
\LinesNumbered
\KwData{Set of objects $\Omega$, number of clusters $k$}
\KwResult{Set of medoids $\Omega^*$ = \{$O_1, O_2, ..., O_k$\}}
\Begin{
$\Omega^* \leftarrow \Omega_k$: start with $K$ objects randomly selected from $\Omega$\;
\Repeat{no change}{
Assign each one of the non-selected objects to the cluster having the most similar medoid\;
Calculate the cost $C_i = (D(\Omega'_k) - D(\Omega^*))$ for each swap of one medoid with another object\;
\If{$(C_i < 0)$}{$\Omega^* \leftarrow \Omega'_k$\;}
}
}
\end{algorithm}

\begin{property}(Termination)
There is only a finite number of possible partitioning of the set of objects $\Omega$ into k groups. 
As we are looking for the partitioning that best minimize the overall distance, we do not go from one partitioning to another only if it improves the distortion. Thus, in each swap the algorithm must choose a \textit{new} partitioning. Consequently, after a number of iterations it would run out of partitioning.
\end{property}
%\subsubsection{Why K-medoids and Not K-means?}
%\textit{K-means} \cite{Frahling_2006} is one of the most used algorithms for clustering. Besides being less sensitive to noise and outliers, we adopt the k-medoids clustering instead of K-means because the latter defines the cluster's centers as fictive points. Thus, in order to detect the subgraph delegates, we have to compute the distance between the subgraphs and the center within the same cluster and consider the closest subgraph to the centeroid as the representative subgraph delegate. Whereas, the k-medoids algorithm requires that the cluster's centroids be real points instead of being fictive. Hence, the cluster's medoids are directly considered as the representative subgraph delegates which prevents performing unnecessary computation needed to detect the delegates with k-means.

\subsubsection{The main algorithm}
We propose \textbf{TRS}, an approach for selecting the \textbf{T}opological \textbf{R}epresentative \textbf{S}ubgraphs. The general algorithm of the approach is described in Algorithm \ref{alg:main_algo}. As previously mentioned, our selection approach follows a two step framework. In the first step, each subgraph is encoded into a topological description-vector using the previously defined topological attributes. In the second step, the set of topological description-vectors are considered for clustering using k-medoids. The selected medoids are considered as the topological representative subgraph-delegates. 

\begin{algorithm}
\label{alg:main_algo}
\caption{\textsc{TRS}}
%\DontPrintSemicolon
\SetAlgoLined
\LinesNumbered
\KwData{Frequent subgraphs $\Omega$, number of representatives $k$}
\KwResult{Topological representative subgraphs $\Omega^*$ = \{$g_1, g_2, ..., g_k$\}}
\Begin{
$M \leftarrow \cup_{i=1}^{|\Omega|} V_i$: each subgraph $g\in\Omega$ is encoded into a topological description vector $V$ using the topological attributes\;
$\Omega^* \leftarrow $K-medoids$(M, k)$\;
}
\end{algorithm}
\begin{property}(Termination)
Since k-medoids terminates, TRS should terminate too.
\end{property}
\section{Experimental analysis}\label{sec:experimental}
\subsection{Datasets}

To experimentally evaluate our approach, we use different types of graph datasets: protein 3D-structures and chemical compounds. Table \ref{tab:datasets} summarizes the characteristics of the four datasets: dataset, $|G|$, Avg.$|V|$, Avg.$|E|$ and $ \mid\Omega\mid$ correspond respectively to the name of the corresponding protein family or chemical compound dataset, number of graph, average number of nodes, average number of edges and number of frequent subgraphs obtained from each dataset. 
%\begin{table*}[!h]
%\centering
%\caption{Benchmark datasets\label{tab:datasets}}
%\begin{tabular}{|c|c|c|c|c|c|c|c|c|}\hline
%Dataset & |$G$| & Pos. & Neg. & Moy.|V| & Moy.|E| & Max.|V| & Max.|E| \\ \hline 
%G-proteins & 66 & 33 & 33 & 246 & 971 & 897 & 3544 \\ \hline
%C1 set domains & 76 & 38  & 38 & 238  & 928  & 768  & 2962 \\ \hline
%Enzymes & 664 & - & - & 358 & 910 & .. & .. \\ \hline
%AIDS antiviral screen & 43850 & - & - & 28 & 30 & .. & .. \\ \hline
%\end{tabular} 
%\end{table*}
\begin{table}[!h]
\centering
\caption{Benchmark datasets\label{tab:datasets}}
\begin{tabular}{ccccccc}\hline
\textbf{Dataset} & \textbf{$\mid$G$\mid$} & \textbf{Moy.$\mid$V$\mid$} & \textbf{Moy.$\mid$E$\mid$} & $ \mid\Omega\mid$\\ \hline 
G-proteins & 66 & 246 & 971	&	114792 \\ \hline
C1 set domains & 76 & 238  & 928	&	258371 \\ \hline
Enzymes & 664 & 358 & 910	&	253404	\\ \hline
AIDS antiviral screen & 43850 & 28 & 30	&	6749 \\ \hline
\end{tabular}
\end{table}

The first two datasets were previously used in \cite{Fei_2010} and \cite{Yan_2008}. Both datasets will be used to evaluate the quality of the selected subgraphs. In fact, each dataset is composed of two groups of protein 3D-structures equally divided between positive and negative samples. Positive proteins are sampled from a selected protein family, namely G-proteins and C1 set domains, whereas negative proteins are randomly sampled from the Protein Data Bank \cite{Berman_2000}. 
%G-proteins are also known as guanine nucleotide-binding proteins. These proteins are mainly involved in transmitting chemical signals originating from outside a cell into the inside of it. They regulate metabolic enzymes, ion channels, transporter, and other parts of the cell machinery, con-trolling transcription, motility, contractility, and secretion, which in turn regulate diverse systemic functions such as embryonic development, learning and memory, and homeostasis. The C1-set domains composing the second dataset are immunoglobulin-like domains, similar in structure and sequence. They resemble the antibody constant domains. They are mostly found in molecules involved in the immune system, in the major histocompatibility complex class I and II complex molecules, and in various T-cell receptors. 
The two other datasets are used to evaluate the runtime and the distribution of subgraphs according to their sizes. The dataset of Enzymes, previously used in \cite{Dobson_2003} and \cite{Marisa_2010}, is composed of 664 proteins. 
%Enzymes act as biological catalysts. They are large biological molecules responsible for the thousands of chemical interconversions that sustain life. 
The last dataset shows a set of antiviral screen data (AIDS). It contains the activity test information of 43850 chemical compounds. This dataset was previously used in many studies such as \cite{Chen_2008} and is publicly available on the website of the Developmental Therapeutics Program \footnotemark[1]\footnotetext[1]{http://dtp.nci.nih.gov/docs/aids/aids$\_$data.html}.

\subsection{Protocol and settings}
\textbf{Graph Building:}
For chemical compounds, each atom is represented by a node and labeled with the atom type (Hydrogen (H), Carbon (C), etc.). An edge exists between two nodes if there exists a chemical bond between their corresponding atoms.
For protein 3D-structures, each protein is parsed into a graph of amino acids. Each node represents an amino acid residue and is labeled with its amino acid type. Two nodes $u$ and $v$ are linked by an edge $e(u, v) = 1$ if the euclidean distance between their two $C_\alpha$ atoms $\Delta(C_\alpha(u), C_\alpha(v))$ is below a threshold distance $\delta$. We use $\delta = 7$\AA . Formally:
\begin{center}
$e(u,v)= \begin{cases}1, \emph{  }if\emph{  } \Delta(C_\alpha(u), C_\alpha(v))\leq \delta
\\
0, \emph{  }otherwise \end{cases}$
\end{center}

\textbf{Frequent subgraph mining:}
We use the state-of-the-art method of frequent subgraph discovery \textsc{gSpan} \cite{Yan_2002} to find the frequent subgraphs in each dataset. We tried different minimum frequency threshold in order to obtain a reasonable number of frequent subgraphs from each dataset. The retained minimum frequency threshold are 30\% for G-proteins and C1 set domains, 10\% for Enzymes, and 5\% for AIDS antiviral screen dataset. %Table \ref{tab:nb_subgraphs} shows the number of frequents subgraphs obtained from each dataset.
%\begin{table}[!h]
%\centering
%\caption{Number of frequent subgraphs ($\Omega$) extracted from each dataset\label{tab:nb_subgraphs}}
%\begin{tabular}{|c|c|}\hline 
%\textbf{Dataset} & $ \mid\Omega\mid$\\ \hline 
%G-proteins & 114792  \\ \hline
%C1 set domains & 258371  \\ \hline
%Enzymes & 253404 \\ \hline
%Sida & 6749 \\ \hline
%\end{tabular} 
%\end{table}

\textbf{Representative subgraph selection:}
Both selection framework, $i.e.$ the na\"ive approach and TRS, were implemented in R. 

\textbf{Subgraph encoding:}
To measure the quality of subgraphs, each one of them is encoded into a binary vector by denoting 1 or 0, the presence or the absence of the subgraph in each graph in the dataset. The quality of the selected subgraphs is measured over their encoding vectors.

\section{Results and discussion}\label{sec:results}
\subsection{Empirical Results}
As previously mentioned, we first evaluate our approach over the classification datasets G-proteins and C1 set domains. We measure the quality of the selected subgraphs using the information gain which is one of the most popular interestingness measures in data mining. Given a set of training examples $\Omega$ and an attribute $att$. The information gain of $att$ is computed using the following formulas:
\begin{center}
$Information Gain(\Omega, att) = Entropy(\Omega) - Enropy(\Omega|att)$
\end{center}
such that $Entropy(\Omega)$ is calculated as follows:
\begin{center}
$Entropy(\Omega) = - \Sigma_{i=1}^{\mid\Omega\mid}p(x_i) log p(x_i)$
\end{center}
where $p(x_i)$ is the probability of getting the $x_i$ value when randomly selecting an example from the set.

The information gain is measured over all the frequent subgraphs then over the subgraphs selected by TRS and those selected by the na\"ive approach using different number of representatives. The information gain value obtained over all the frequent subgraphs is considered as standard value for comparison. Table \ref{tab:inf_gain} shows the obtained results.
\begin{table}[!h]
\centering
\caption{Comparison of average information gain of the topological representative subgraphs (TRS) with those selected by the na\"ive approach (NA) and the initial set of all frequent subgraphs (FSG).\label{tab:inf_gain}}
\begin{tabular}{ccccccc}\hline
%\cline{2-5}
& \multicolumn{2}{c}{\textbf{G-proteins}} & \multicolumn{2}{c}{\textbf{C1 set domains}} \\\hline \hline
 \multicolumn{1}{c}{\textbf{FSG}} & \multicolumn{2}{c}{0.216} &  \multicolumn{2}{c}{0.148} \\ \hline \hline 
 \multicolumn{1}{c}{\textbf{\# representatives}} & \textbf{NA} & \textbf{TRS} & \textbf{NA} & \textbf{TRS} \\ \hline 
 \multicolumn{1}{c}{50}  & 0.104  & 0.324 & 0.068 & 0.254 \\ \hline 
 \multicolumn{1}{c}{100} & 0.092  & 0.342 & 0.061 & 0.285 \\ \hline
 \multicolumn{1}{c}{200} & 0.096  & 0.343 & 0.044 & 0.273 \\ \hline
 \multicolumn{1}{c}{300} & 0.097  & 0.347 & 0.058 & 0.267 \\ \hline 
 \multicolumn{1}{c}{400} & 0.094  & 0.339 & 0.051 & 0.276 \\ \hline 
 \multicolumn{1}{c}{500} & 0.090  & 0.348 & 0.052 & 0.269 \\ \hline 
 \multicolumn{1}{c}{600} & 0.096  & 0.340 & 0.054 & 0.267 \\ \hline
 \multicolumn{1}{c}{700} & 0.097  & 0.343 & 0.055 & 0.272 \\ \hline
 \multicolumn{1}{c}{800} & 0.098  & 0.352 & 0.054 & 0.274 \\ \hline
 \multicolumn{1}{c}{900} & 0.094  & 0.358 & 0.054 & 0.276 \\ \hline
 \multicolumn{1}{c}{1000} & 0.094  & 0.353 & 0.056 & 0.276 \\\hline
 \textbf{Average} & $0.095_{-0.005}^{+0.008}$ & $0.344_{-0.020}^{+0.013}$ & $0.055_{-0.011}^{+0.012}$ & $0.271_{-0.017}^{+0.013}$ \\ \hline
 
\end{tabular} 
\end{table}

Table \ref{tab:inf_gain} shows that TRS is able to select a subset of subgraphs that are more informative than those selected by the na\"ive approach and the initial frequent subgraphs. Whereas, the quality of the subsets of representative subgraphs selected by the na\"ive approach did not even reach the information gain value of the whole set of frequent subgraphs. Both previous interpretations goes with all the used numbers of representatives. This proves the reliability of our selection approach and shows that using the topological attributes for description is more efficient than using the occurrence information. It enables k-medoids to better detects similarity relations between subgraphs and thus to select a subset of representatives that are most informative.% It also shows that the topological description allowed detecting relations between subgraphs, that are ignored by the na\"ive approach. 
\subsection{Size-based distribution of patterns}
\begin{figure}
	\centering
	\includegraphics[width=0.45\textwidth]{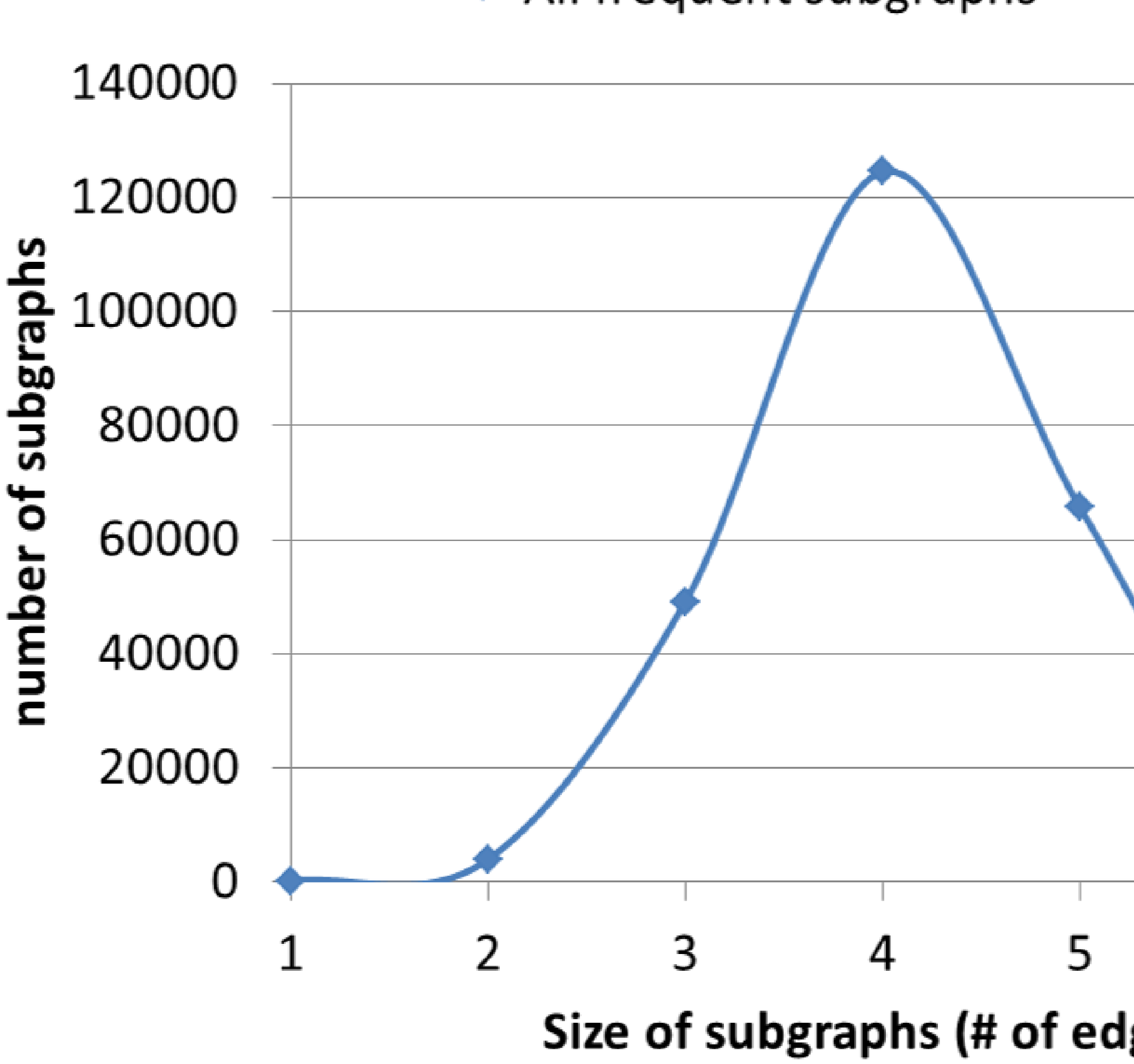}\includegraphics[width=0.45\textwidth]{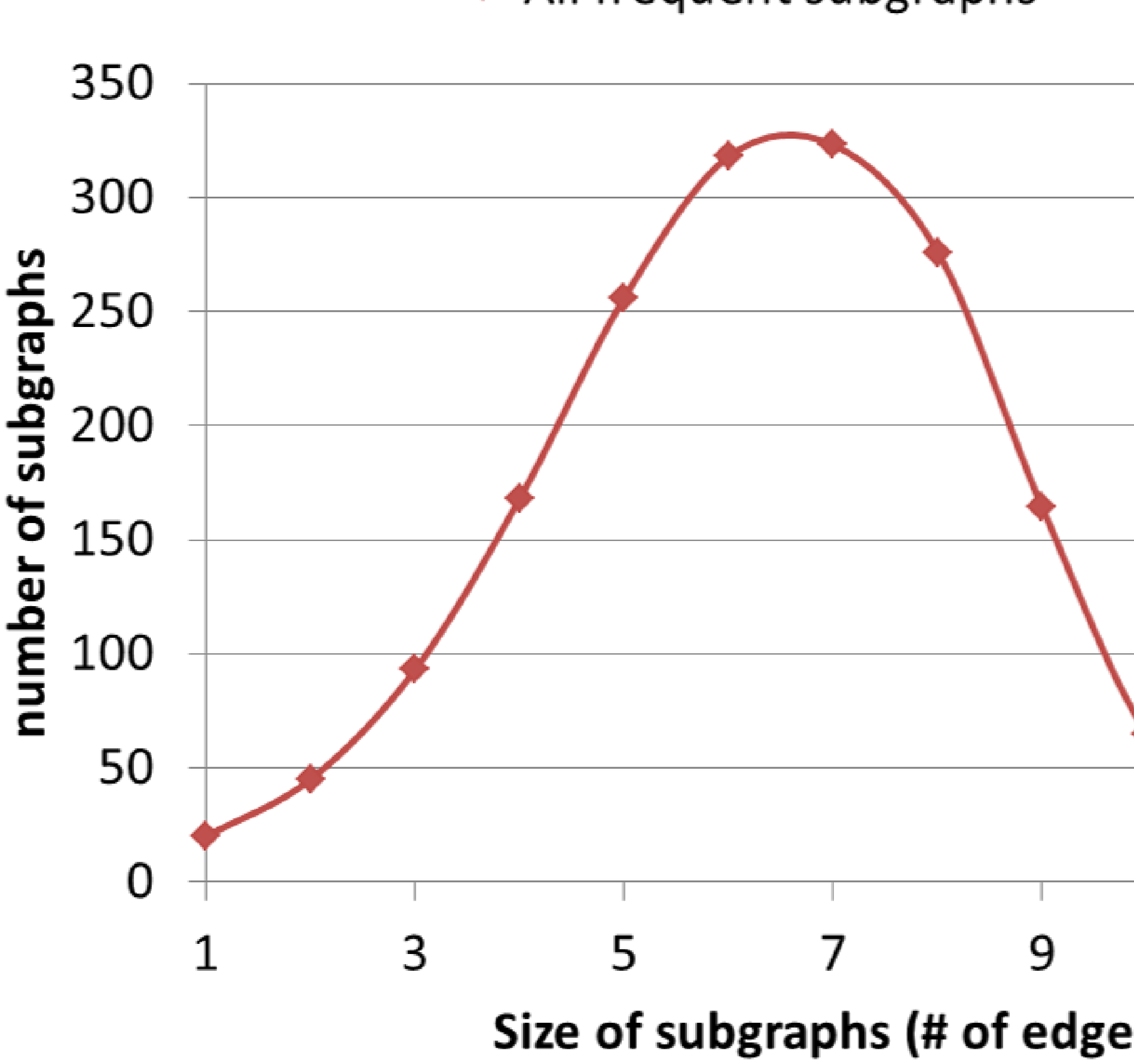}
	   \caption{Distribution of subgraphs by size for the Enzymes (left) and the AIDS antiviral screen (right) datasets.} \label{fig:initial_distribution_enzymes}
\end{figure}

\begin{figure}
	\centering
	\includegraphics[width=0.45\textwidth]{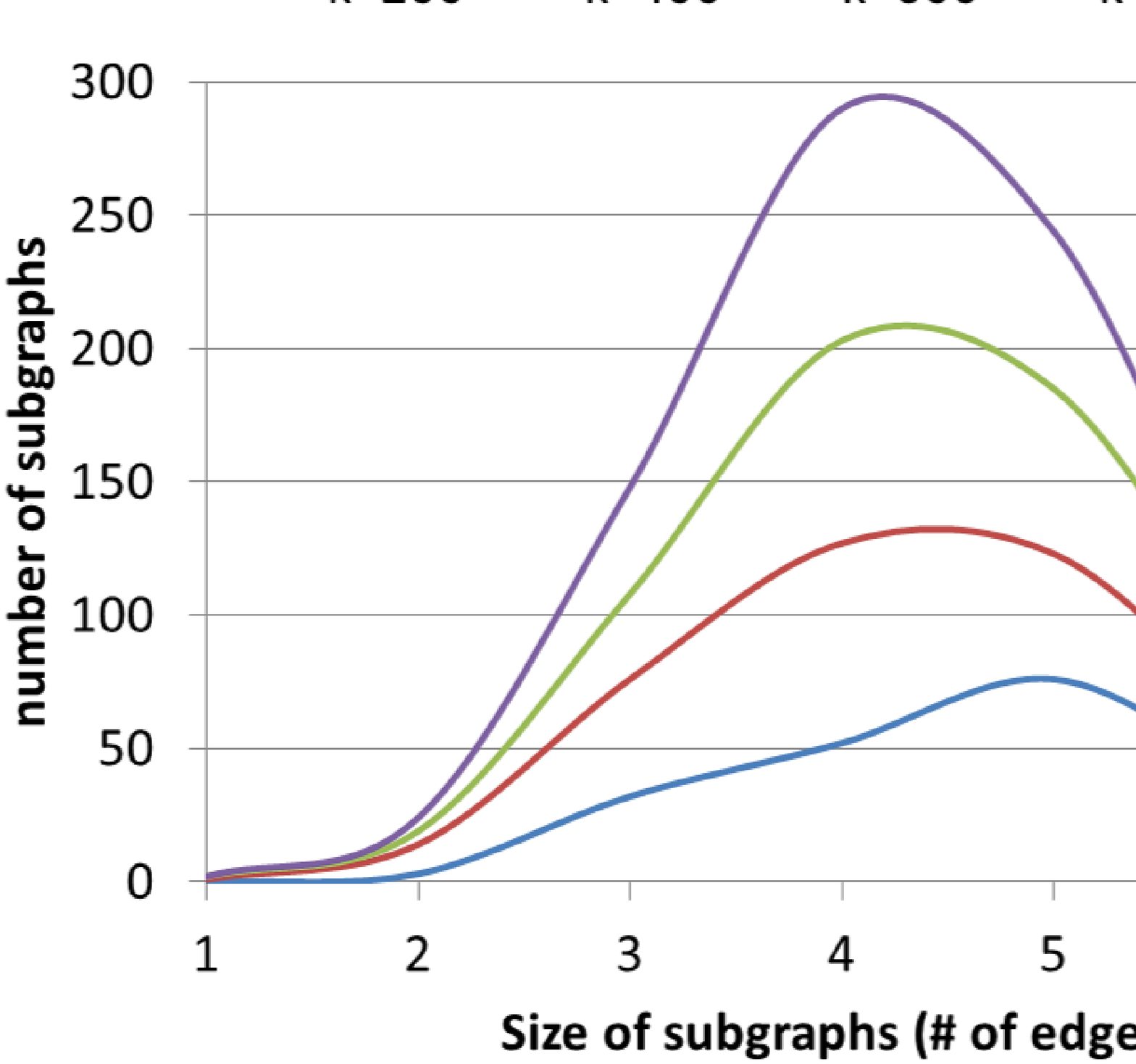}\includegraphics[width=0.45\textwidth]{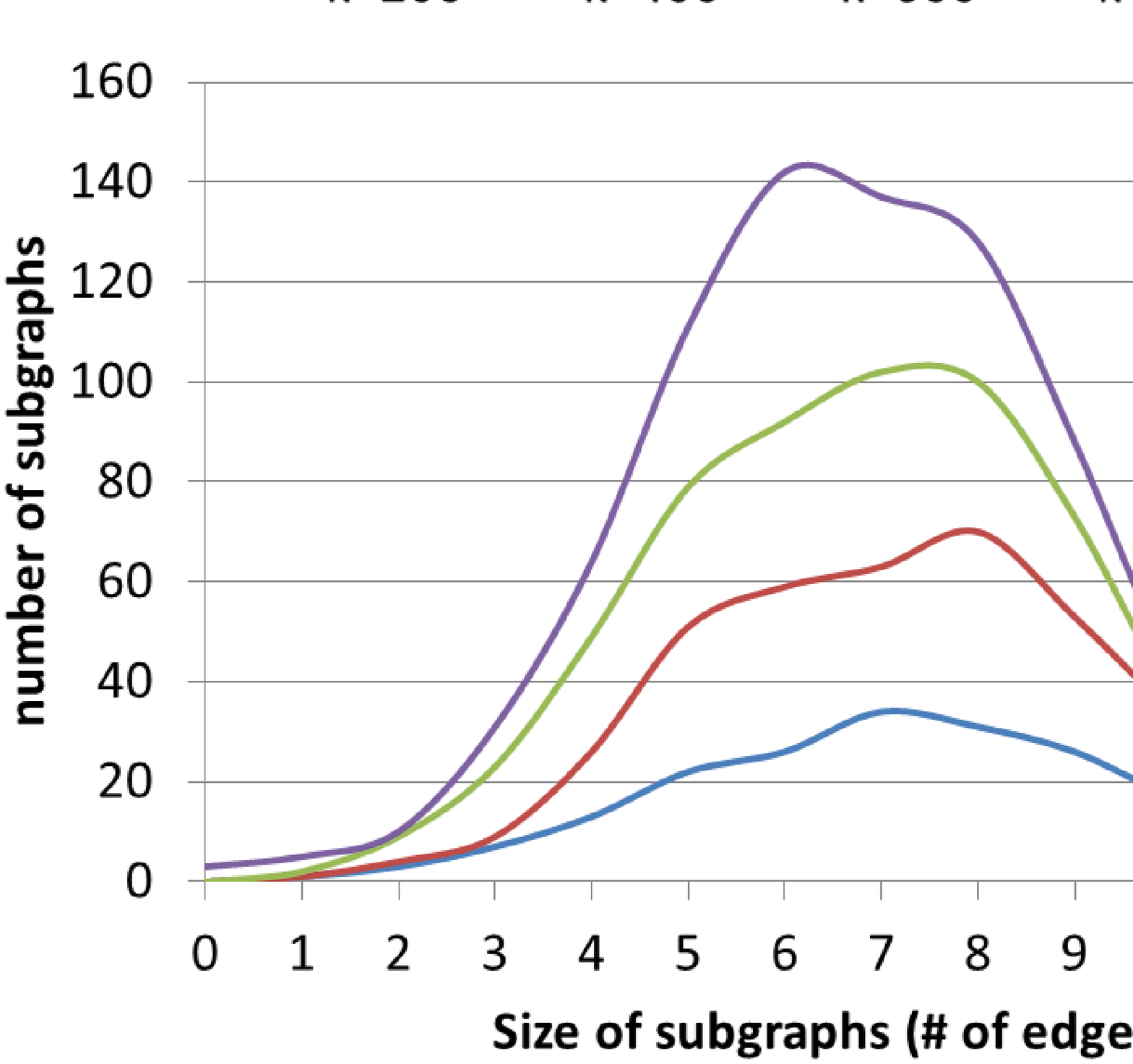}
	   \caption{Distribution of subgraphs by size for the Enzymes (left) and the AIDS antiviral screen (right) datasets using different values of $k$(number of representatives).} \label{fig:distribution_enzymes}
\end{figure}

In this section, we study the distribution of subgraphs based on their size (number of edges). We try to check which sizes are more concerned by the selection. Figure \ref{fig:initial_distribution_enzymes} draws the distribution of the original set of frequent subgraphs over Enzymes and AIDS antiviral screen datasets. For both datasets, we notice a high concentration of the number of frequent subgraphs in the center especially with the Enzymes dataset.%, ranging from three to five for the latter, and from four to nine for the AIDS antiviral screen dataset. 
These concentration zones presents high level of redundancy and must be the most concerned by the selection.
%\begin{figure}
%	\centering
%	
%	   \caption{Distribution of subgraphs by size for  dataset.} \label{fig:initial_distribution_aids}
%\end{figure}
Figure \ref{fig:distribution_enzymes} draws the distribution of the topological representative subgraphs with different number of clusters $k$. The downward tendency of TRS using lower values of $k$ and with respect to the original set of frequent subgraphs is very clear. In fact, TRS leans towards cutting off the peaks and flattening the curves with lower value of $k$. Another interesting observation is that the curves are flattened in the regions of small subgraphs as well as in those of big subgraphs. This demonstrates the effectiveness of TRS with both small and big subgraphs.
%\begin{figure}
%	\centering
%	
%	   \caption{Distribution of subgraphs by size for the AIDS antiviral screen dataset using different values of $k$(number of representatives).} \label{fig:distribution_aids}
%\end{figure}

\subsection{Runtime analysis}
In this section, we study the runtime of our algorithm compared to that of the na\"ive approach on three levels: in terms of variation of number of clusters, numbers of frequent subgraphs, and number of graphs. It is worth mentioning that here we only compare the clustering runtime and we omit the time of the encoding of subgraphs since it does not change along the experiments and only counts few seconds. Besides, it depends on the selected attributes for TRS.
\subsubsection{Scalability to higher number of clusters}
\begin{figure}
	\centering
	\includegraphics[width=0.45\textwidth]{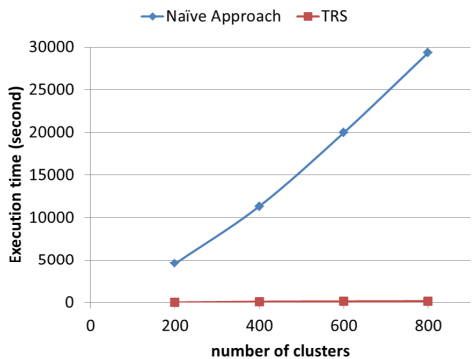}
	   \caption{Runtime of clustering for TRS and the na\"ive approach with different number of clusters ($k$).} \label{fig:runtime}
\end{figure}

We study the effect of varying the number of clusters $k$ on the runtime of clustering for both TRS and the na\"ive approach. We select the representative subgraphs among the frequent ones previously extracted from the AIDS antiviral screen dataset. Figure \ref{fig:runtime} illustrates the evolution of runtime using different values of $k$ (number of clusters) ranging from 200 to 800 with a step-size of 200.

Figure \ref{fig:runtime} shows a huge difference in execution time between the two approaches. In fact, for 200 clusters, the na\"ive approach consumed more than one and half hour to finish the clustering, whereas TRS needed only few seconds. This difference becomes much bigger with higher values of $k$. As the number of clusters increases, the execution time of the na\"ive approach exponentially increases as well. Yet, the clustering time in TRS did not increase significantly and almost stays steady with higher values of $k$. Since the clustering is combinatorial and considers each possible pair of subgraphs for comparison, the smaller the description of the subgraphs is, the faster the clustering would be. Consequently, the huge gain in execution time is basically due to the small and fixed size of the topological description-vectors used in TRS compared to the context description-vectors in the na\"ive approach. 
\subsubsection{Scalability to higher number of subgraphs}
\begin{figure}
	\centering
	\includegraphics[width=0.45\textwidth]{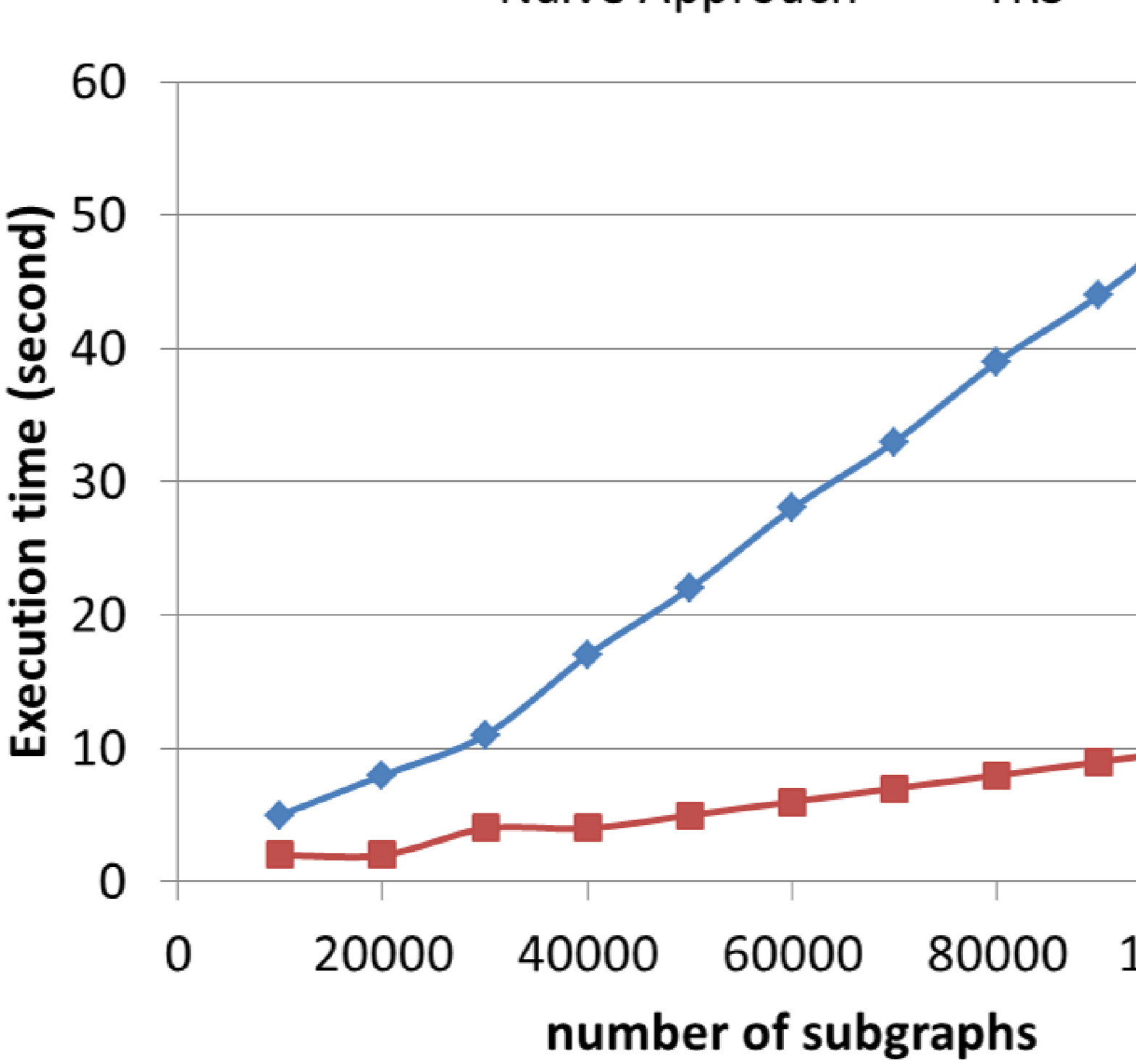}	\includegraphics[width=0.45\textwidth]{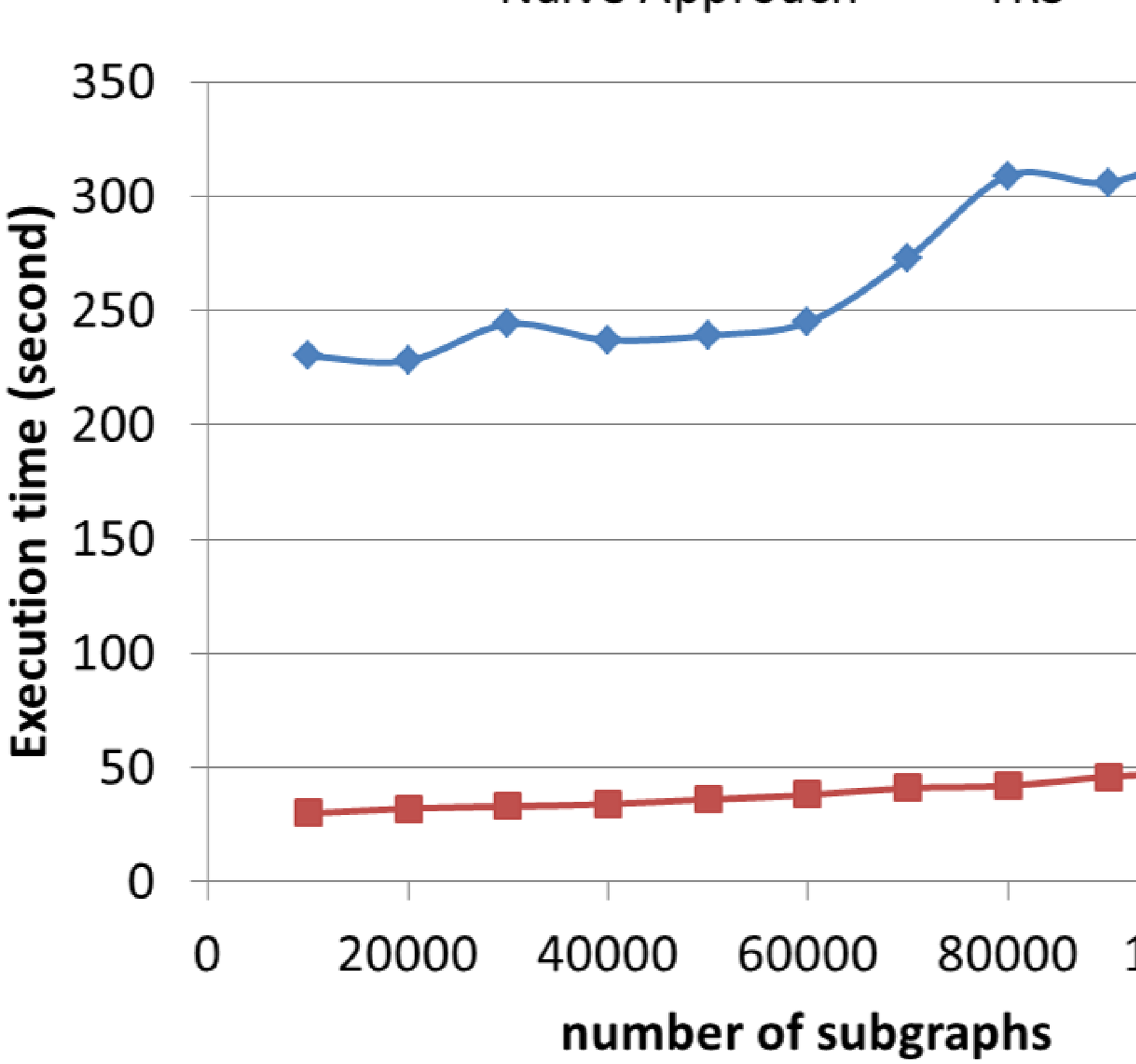}
	   \caption{Runtime of clustering for TRS and the na\"ive approach to select 100 (left) then 500 (roght) representatives among different numbers of subgraphs.} \label{fig:100_subgraphs}
\end{figure}

Here, we study the effect of varying the number of frequent subgraphs on clustering runtime for both TRS and the na\"ive approach. We select the representative subgraphs among different sets of frequent subgraphs ranging from 10000 to 100000 with a step size of 10000. The input subgraphs were randomly selected among the frequent subgraphs previously extracted from the C1 set domains dataset. Figure \ref{fig:100_subgraphs} illustrates the evolution of runtime with higher number of subgraphs, for 100 and 500 clusters.
%\begin{figure}
%	\centering
%
%	   \caption{Runtime of clustering for TRS and the na\"ive approach to select 500 representatives among different number of subgraphs.} \label{fig:500_subgraphs}
%\end{figure}
As shown in the figure, TRS only takes few seconds to select the representative subgraphs, whereas, the na\"ive approach takes clearly much more time. Increasing the number of subgraphs does no affect the runtime of TRS as much as it does with the na\"ive approach. This shows that TRS is more scalable than the na\"ive approach to higher numbers of subgraphs.
\subsubsection{Scalability to higher number of graphs}
\begin{figure}
	\centering
	\includegraphics[width=0.45\textwidth]{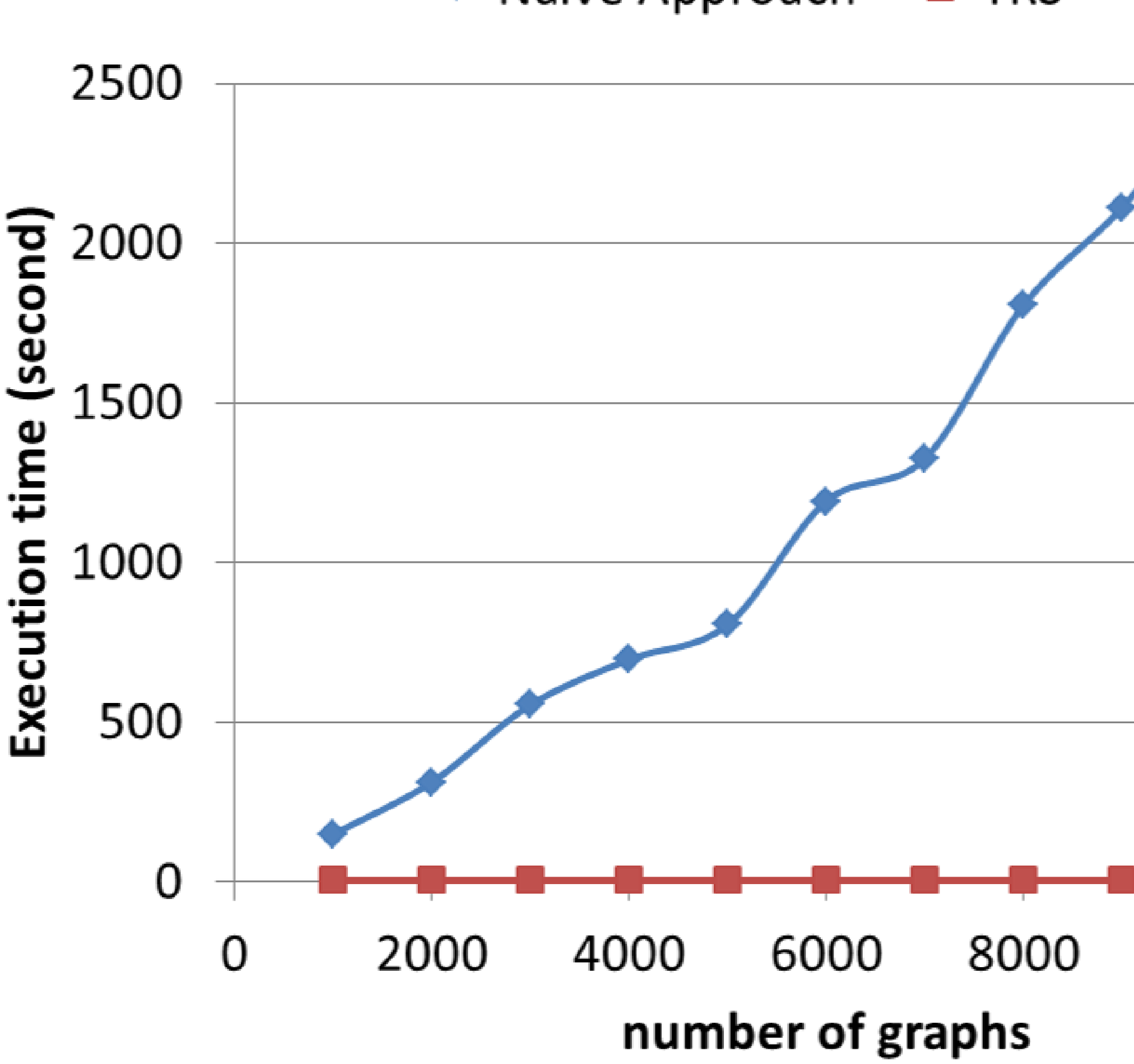}	\includegraphics[width=0.45\textwidth]{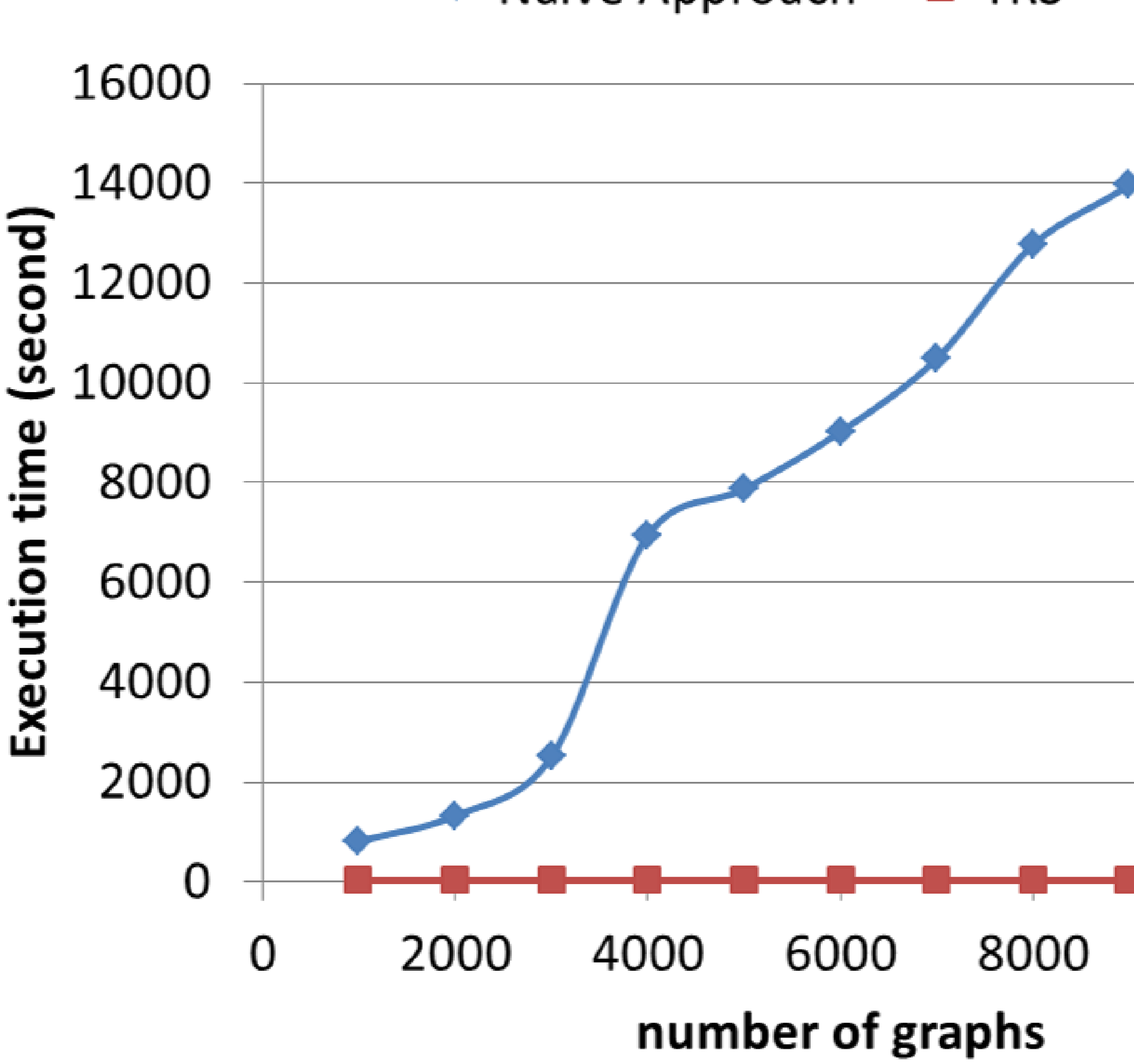}
	   \caption{Runtime of clustering for TRS and the na\"ive approach to select 100 (left) then 500 (right) representatives among 10000 subgraphs whith variation of the number of graphs.} \label{fig:100_graphs}
\end{figure}

In real-world applications, the size of graph databases is usually very high. We study the effect of varying the number of graphs on the runtime of both TRS and the na\"ive approach. We fix the number of subgraphs to 10000, and we synthetically manipulate the list of occurrences of each frequent subgraph and replace it by a random list of random occurrences between 0 and a considered number of graphs. The considered numbers of graphs are between 1000 and 10000, with a step size of 1000. Figure \ref{fig:100_graphs} illustrates the evolution of runtime with higher number of graphs, respectively for 100 and 500 clusters.
%\begin{figure}
%	\centering
%
%	   \caption{Runtime of clustering for TRS and the na\"ive approach to select 500 representatives among 10000 subgraphs whith variation of the number of graphs.} \label{fig:500_graphs}
%\end{figure}

As the na\"ive approach uses the occurrence information to construct the context description-vectors, this makes it highly affected by the increasing of the size of the database. Figure \ref{fig:100_graphs} shows that the runtime of the na\"ive approach increases exponentially with higher numbers of graphs. Whereas, the runtime of TRS corresponds only to few seconds and remains stable no matter what the size of the database is. This shows that TRS is scalable and more robust in real world-applications that usually deals with huge amounts of data.
\section{Conclusion}
We proposed a novel approach that mines a subset of topological representative subgraphs among the frequent ones. Instead of exact and approximate structural similarity our approach follows a more meaningful selection strategy, that helps on both selecting a subset of topologically irredundant  and informative subgraph-delegates, and detecting hidden relations between subgraphs that are ignored by current selection approaches. This approach can be easily extended using any user defined structural or topological attributes. Besides graph databases, it can handle other cases such as the problem of subgraph selection in single graph. Empirical studies on real and synthetic graph datasets show that our approach is fast and scalable.
In many application, the user may not be able to define a specific number of clusters. A promising future direction could be to remove the \textit{k} constraint. This can be simply done using a parameter free clustering algorithm such as Medoids-shifts \cite{Sheikh_2007}. As in real-world application the number of subgraphs can be exponential, it would be also interesting to make the approach even more scalable using a parallel clustering algorithm such as CAPEK \cite{Gamblin_2010}.
%Since the proposed method is a filter approach, a promising future direction could be to find a way to integrate the selection within the extraction process in order to directly mine the representative subgraphs from data.

%
%\bibliographystyle{elsarticle-num}
%\bibliography{Thesis}

%% Authors are advised to submit their bibtex database files. They are
%% requested to list a bibtex style file in the manuscript if they do
%% not want to use elsarticle-num.bst.

%% References without bibTeX database:

\end{document}